\documentclass[10pt,tightenlines,eqsecnum,floats,aps,amsmath,amssymb,nofootinbib,superscriptaddress,prd,showpacs]{revtex4}

\usepackage{amsmath,amsthm,latexsym,amssymb,amsfonts}
\usepackage{enumerate}
\usepackage{graphicx}
\usepackage{calc}
\usepackage{color}
\usepackage{vmargin}

\newcommand{\Hil}{{\mathcal H}}

\newcommand{\Hphy}{{\Hil}_{\rm phys}}
\newcommand{\Cgr}{{C}^{\rm gr}}

\begin{document}

\title{Gravity Quantized: Loop Quantum Gravity with a Scalar Field} 

\author{Marcin \surname{Domaga{\l}a}}\email{marcin.domagala@fuw.edu.pl}\affiliation{Instytut Fizyki Teoretycznej, Uniwersytet Warszawski, ul. Ho{\.z}a 69, 00-681 Warszawa (Warsaw), Polska (Poland)}

\author{Kristina  \surname{Giesel}}\email{kgiesel@phys.lsu.edu}\affiliation{Excellence Cluster Universe, Technische Universit\"at M\"unchen, Boltzmannstr. 2, 85748 Garching, Germany}\affiliation{Department of Physics and Astronomy, Louisiana State University, Baton Rouge LA 7080 3, United States}

\author{Wojciech \surname{Kami\'nski}}\email{wojciech.kaminski@fuw.edu.pl}\affiliation{Instytut Fizyki Teoretycznej, Uniwersytet Warszawski, ul. Ho{\.z}a 69, 00-681 Warszawa (Warsaw), Polska (Poland)}

\author{Jerzy \surname{Lewandowski}}\email{jerzy.lewandowski@fuw.edu.pl}\affiliation{Instytut Fizyki Teoretycznej, Uniwersytet Warszawski, ul. Ho{\.z}a 69, 00-681 Warszawa (Warsaw), Polska (Poland)}\affiliation{Institute for Gravitation and the Cosmos, Physics Department, Penn State, University Park, PA 16802, U.S.A.}

\date{15 September. 2010}

\begin{abstract} \noindent{\bf Abstract\ }
...''but we do not have quantum gravity.'' This phrase is often used when  analysis of a physical problem enters the regime in which quantum gravity effects should be taken into account. In fact, there are several models of the gravitational field coupled to  (scalar) fields for which the quantization procedure can be completed using loop quantum gravity techniques. The model we present in this paper consist of the gravitational field coupled to a scalar field. The result has similar  structure to the loop quantum cosmology  models, except for that it  involves all the local degrees of freedom of the gravitational field{ because no symmetry reduction has been performed at the classical level}.                
\end{abstract}


\maketitle

\def\be{\begin{equation}}
\def\ee{\end{equation}}
\def\ben{\begin{equation*}}
\def\een{\end{equation*}}
\def\ba{\begin{eqnarray}}
\def\ea{\end{eqnarray}}
\def\ban{\begin{eqnarray*}}
\def\ean{\end{eqnarray*}}

\def\lp{{\ell}_{\rm Pl}}
\def\g{\gamma}

\section{Introduction} 

The recent advances in  loop quantum gravity (LQG)  \cite{AshTat, AshLew1, Thiemm2, Rov2}, strongly suggest that the goal of constructing a  candidate for quantum theory of gravity and the Standard Model is within reach. Remarkably, that goal can be  addressed within the  canonical formulation of  the original Einstein's general relativity in four dimensional spacetime.   A way to define 'physical' dynamics in a background independent theory, where spacetime diffeomorphisms are treated as a gauge symmetry, is the framework of relational Dirac observables (often also called ``partial'' observables \cite{Rov1},\cite{Ditt1,Ditt2},\cite{ThiemannRed} section I.2 of \cite{Thiemm2}). The main idea is, that part of the fields adopt the role of a dynamically coupled observer, with respect to which the physics of the remaining degrees of freedom in the system is formulated. In this framework  the emergence of the dynamics, time and space can be explained as an effect of the relations between the fields. As far as technical issues of a corresponding quantum theory are concerned, the most  powerful example of the relational observables framework is the deparametrization technique \cite{KijSmoGor,IacKijMag, BroKuch, Thiemm3}. This allows to map canonical General Relativity into a theory with a (true) non-vanishing Hamiltonian, that is independent of the (emergent) time provided by the observer fields. All this can be achieved at the classical level, the framework of  Loop Quantum Gravity (LQG) itself, provides  then the tools of the quantum theory like quantum states, the Hilbert spaces, quantum operators of the geometry and fields  and well defined  quantum operators for the classical constraints of General Relativity (see \cite{Thiemm2},\cite{AshLew1} and references therein). The combination of LQG with the relational observables and deparametrization framework makes  it possible   to construct general relativistic quantum models. Applying LQG techniques to perform the  quantization step has the consequence that the quantum fields of the Standard Model have to be reintroduced within the scheme of LQG. This is due to the reason that the standard  quantum field theory (QFT) defined on the Minkowski (or even ADS) background is incompatible with quantization approach used in LQG. Therefore, the  resulting quantum theory of gravity cannot be just coupled to the Standard Model in it's present form.  The formulation of the full Standard Model within LQG will require some work. For this reason, we proceed step by step, increasing { gradually} the level of complexity. The first step was constructing various cosmological models by analogy  with LQG by performing a symmetry reduction already at the classical level. They give rise to loop quantum cosmology \cite{Boj1, Boj2, Boj3, Boj4, AshBojLew, AshPawSin1, AshPawSin2} (LQC). We have learned from them a lot about qualitative properties of quantum spacetime and its quantum dynamics \cite{KamLewPaw1,KamLewPaw2}. That knowledge is very useful in performing the second step, that is introducing quantum models with the full set of the local gravitational degrees of freedom. The first  quantum model of the full, four dimensional  theory of gravity was obtained by applying LQG techniques \cite{GieThiemm} to the Brown-Kuchar model of gravity coupled to dust  \cite{BroKuch}. In the current paper we apply LQG to the model introduced by Rovelli and Smolin \cite{RS} whose  classical canonical structure was studied in detail by Kuchar-Romano \cite{KuchRom1}. This is a model of gravity coupled to a massless scalar field. Our goal is to complete the construction of the quantum model with the tools of LQG. In the firs part of the paper (Sections I-II)  we introduce the model, study the  structure of the space of solutions to the quantum constraints, and the Dirac observables, assuming only suitable Hilbert products and operators exist. The result of this part is a list of mathematical elements necessary and sufficient for the model to exist. In the second part (Section III) we apply the framework of LQG. We show it provides the necessary Hilbert spaces and operators, and complete  the construction of the model.    

\section{Canonical gravity coupled to a classical scalar field}

\subsection{The standard approach}
The point of our interest in this paper is gravity coupled to  a scalar field. We are considering a metric tensor field $q_{ab}$ and a scalar field $\phi$ on a 3-manifold $M$ (the space). The conjugate momenta are denoted respectively by  $p^{ab}$ and $\pi$. The only non-vanishing Poisson brackets among them are 
\be
\{q_{ab}(x),p^{cd}(y)\}\ =\ \delta(x,y)\delta_{(a}^c\delta_{b)}^d,\ \ \ \{\phi(x),\pi(y)\}\ =\ \delta(x,y).
\ee
The  intrinsic and extrinsic geometry of $M$ (as $M$ being the Cauchy surface of 4-dimensional  spacetime) is described by the first pair of  canonically conjugate variables   $(q_{ab},p^{ab})$. The field $q_{ab}$ defines the intrinsic Riemann geometry of $M$ whereas $p^{ab}$ contains the information about the extrinsic curvature of $M$ imbedded in the spacetime.

The variables  $(q_{ab},p^{ab})$ are known from the standard canonical formulation of  gravity usually called ADM formalism \cite{ADM} (see also chapter 10 and  appendix E of  \cite{Wald}). But one can use any other variables in this part of our paper (Section I-II). In Section III we will apply  loop quantum gravity (LQG), and therein we will be using the Ashtekar-Barbero   variables $(A^i_a, P_i^a)$, $i=1,2,3$ (and the notation of \cite{AshLew1}). They are also canonically conjugate to each other, and the only non-vanishing Poisson bracket is
\be 
\{A^i_a(x),\ P_j^b(y)\}\ =\ \delta(x,y)\delta_a^b\delta^i_j. 
\ee 
The intrinsic and extrinsic geometry of $M$ can be recovered out of them, as they are defined by the orthonormal coframe $e_a^i$, the corresponding connection 1-form $\Gamma^i_a$, the extrinsic curvature 1-form $K_a^i$ and a fixed Barbero-Immirzi parameter $\gamma$ (for its value see\cite{DomLew,Meis,Barbero}), namely
\be 
A^i_a\ =\ \Gamma^i_a+\gamma K^i_a, \ \ \ \ P^c_i\ =\ \frac{1}{16\pi G\gamma}e^j_ae^k_b\eta^{abc}\epsilon_{ijk}
\ee 
where $\eta^{123}=1=\epsilon_{123}$ and $\eta^{abc}$, $\epsilon_{abc}$ are completely antisymmetric. 
 
The fields $(A_a^i,P_i^a)$ set an su(2) valued 1-form, and, respectively, su(2)$^*$ valued vector density 
\be 
A =\ A^i_a(x)\tau_i\otimes dx^a, \ \ \ P\ =\ P^a_i(x)\tau^i\otimes\frac{\partial}{\partial x^a} 
\ee
where $x^a$ are local coordinates in $M$, $\tau_1,\tau_2,\tau_3\in$ su(2) is a basis such that 
\ben
\eta(\tau_i,\tau_j)\ :=\ -2{\rm Tr}(\tau_i\tau_j)=\delta_{ij},
\een
and $\tau^1,\tau^2,\tau^3$ is the dual basis.

Einstein's theory of gravity is subject to constraints. In the standard ADM approach we have two constraints, namely the vector constraint generating the diffeomorphisms of $M$ and the scalar constraint generating dynamics, that is diffeomorphisms orthogonal to the Cauchy hypersurface $M$:
\be
C_a(x)\ =\  \Cgr_a(x) \ +\ \pi(x)\phi_{,a}(x)\label{vector},
\ee
\be
C(x)\ =\Cgr(x)\ +\ \frac{1}{2}\frac{\pi^2(x)}{\sqrt{q(x)}} +\frac{1}{2}q^{ab}(x)\phi_{,a}(x)\phi_{,b}(x)\sqrt{q(x)} + V(\phi)\sqrt{q(x)},\label{scalar}
\ee
where the terms $\Cgr_a$ and $\Cgr$ involve the gravitational field variables  $q_{ab}$ and $p^{ab}$ only.

In LQG, the fields $q_{ab}$ and $p^{ab}$ in the constraints are expressed by the variables $A^i_a$ and $P^a_i$, and  we get an additional constraint - the Gauss constraint generating the ``Yang-Mills''\footnote{Although we do not consider the Yang-Mills theory itself, the Ashtekar-Barbero variables are subject to the gauge transformations known from the Yang-Mills theory.} gauge transformations of the fields $(A,P)$:
\be
G^i_a(x)\ = \partial_aP_i^a + \epsilon_{ij}{}^k A^j_aP^a_k.\label{gauss}
\ee

All the transformations generated by the vector, scalar and the Yang-Mills constraint are gauge transformations, because the constraints are of first class. 

In  Section I and Section II the choice of the variables describing the gravitational part does not matter, so one can either use the ADM variables $(q_{ab},p^{ab})$ and the constraints (\ref{vector}, \ref{scalar}) or, respectively, the Ashtekar-Barbero variables $(A^i_a,P^a_i)$ and,  the constraints (\ref{vector}, \ref{scalar}, \ref{gauss}). In Section III,  the latter choice is necessary, because we will apply LQG. For the sake of the continuity, we will stick to the Ashtekar-Barbero variables, remembering that $q_{ab}$,  $p^{ab}$,$\Cgr$ and  $\Cgr_a$ should be considered as functions of $(A^i_a,P^a_i)$. 

Each choice of the fields $(A^i_a,P^a_i,\phi,\pi)$ defines a point in the phase space $\Gamma$. The solutions to the constraints form a constraint surface. We will also consider separately the phase space of gravitational degrees of freedom denoted by $\Gamma_{\rm gr}$, which by definition is set by the pairs $(A^i_a,P^a_i)$.

By assuming that the vector and the scalar constraints are satisfied
\be 
C(x)=0, \ \ \ \ \ C_a(x)\ =\ 0 \label{zero}
\ee
we can solve the vector constraint in (\ref{vector}) for the  gradient $\phi_{,a}$ obtaining  $\phi_{,a}=-\frac{\Cgr_a}{\pi}$ and  inserting this into the scalar constraint (\ref{scalar}).  What we get, remembering  (\ref{zero}) and solving the scalar constraint for $\pi$, is an expression for $\pi^2$ as a function of the geometry variables $(A_a^i,P^a_i)$ and the potential $V(\phi)$ only,
\be
\pi^2\ =\ \sqrt{q}\left( -\left(C^{GR}+\sqrt{q}V(\phi)\right)\pm \sqrt{\left(C^{GR}+\sqrt{q}V(\phi)\right)^2-q^{ab}C^{GR}_aC^{GR}_b} \right).\label{pm}
\ee 
The ambiguous sign $\pm$  in (\ref{pm}) defines different regions in the phase space $\Gamma$. In particular, only the choice of a plus sign includes
the special case of a  homogenous and isotropic geometry coupled to a scalar field. In the case of the minus sign specialized to cosmological spacetimes, where each vector constraints vanishes identically, the expression for $\pi^2$ above will just yield zero on the righthand side.

\subsection{A deparametrized model}
 
What we have done in the last section is solving the scalar constraint for the scalar field momentum by using the vector constraint. Physically, this corresponds, as will be explained more in detail below, to choose the scalar field $\phi$ as our emergent time with respect to which the dynamics of the observables will be formulated. This calculation provides the relation between the standard real scalar field coupled to gravity, on the one hand, and the model we actually {\it define} below, on the other hand. 

In our paper we will consider a model, that is defined by the vector constraint (\ref{vector}), the Gauss constraint (\ref{gauss}) and the following scalar constraint
\ba
C'(x)\ &=&\ \pi(x) - h(x),\label{newscalar}\\
h\ &:=&\ \sqrt{-\sqrt{q}C^{gr}+\sqrt{q}\sqrt{({C^{gr}})^2-q^{ab}C^{gr}_aC^{gr}_b}}.
\label{hclas}\ea
The scalar constraint $C(x)$ has been rewritten using (\ref{pm}).
That theory is equivalent to the theory defined in the previous subsection in the case of no potential
\be
V(\phi)=0 
\ee
and in the region of the phase space $\Gamma$ such that `$+$' holds in (\ref{pm}) and   
\be
\pi>0.
\ee
Since the potential is set to zero in the model, $\phi$ no longer occurs in the function $h$ and the scalar constraints deparametrizes.
Notice, that in the consequence of the constraints, in that region
\be 
\Cgr \ <\ 0.\label{Cgrclass<0} 
\ee   
The deparametrized scalar constraints, being linear in the scalar field momentum, strongly Poisson commute 
\be
\{C'(x),C'(y)\}\ =\ 0,
\ee
as a consequence of the following identity
\be \label{commclas}
\{h(x),h(y)\}\ =\ 0
\ee
proved in \cite{KuchRom1}.
A Dirac observable is the restriction to the constraint surface of a function $f:\Gamma\rightarrow \mathbb{R}$, such that 
\be 
\{f,C_a(x)\}\ =\ \{f,C'(x)\}\ =\ \{f,G^i_a(x)\} \ =\ 0.
\ee
The vanishing of the first Poisson bracket means, that $f$ is invariant with respect to the action of the local diffeomorphisms (that is all diffeomorphisms generated by the vector fields tangent to $M$), the vanishing of the third Poisson bracket is equivalent to the Yang-Mills gauge invariance of $f$. The vanishing of the second Poisson bracket reads
\be 
\{f,\pi(x)\}\ =\  \{f,h(x)\}.
\ee   

\section{Quantum canonical gravity coupled to a scalar field} 

In this section we introduce a ``formal'' structure of our theory. Our goal, at this point,  is to conclude what mathematical structures (Hilbert spaces, operators etc.) are needed to complete the quantization of the model. How to construct them using LQG will be explained in the next section.

Assuming for the time being, that all  Hilbert spaces and  operators we need exist, and that they have the usual properties, we will now derive: 
\begin{itemize}
 \item  a general solution to the quantum constraints,
 \item  a general quantum Dirac observable, its classical interpretation and its physical evolution,
 \item  the Hilbert product between two solutions.
\end{itemize}

\subsection{Quantum states and quantum fields}
The quantum states are complex valued functions
\be (\phi,A)\ \mapsto\ \Psi(\phi,A), \ee
where $\phi$ and $A$ are the scalar field and the Ashtekar-Barbero connection defined
on $M$ in the previous section (henceforth, we will  write $A$ and $P$ instead of $A^i_a$ and $P^a_i$). 

For a given representation the fields $\phi,\pi,A,P$ give rise to quantum operators 
\ba
\hat{\phi}(x)\Psi(\phi,A)\ &=&\ \phi(x)\Psi(\phi,A), \ \ \ \ \ \ \hat{\pi}(x)\Psi(\phi,A)\ =\ \frac{1}{i}\frac{\delta}{\delta \phi(x)}\Psi(\phi,A) \label{quanphipi}\\
 \hat{A}^j_{b}(x)\Psi(\phi,A)\ &=&\ A^j_{b}(x)\Psi(\phi,A) \ \ \ \ \hat{P}^{b}_j(x)\Psi(\phi,A)\ =\ \frac{1}{i}\frac{\delta}{\delta A^j_{b}(x)}\Psi(\phi,A)\label{quanAP}
\ea 
These elementary operators are needed to define the operators corresponding to the classical constraints and to define
the quantum observables.

\subsection{The quantum constraints and their solutions}

We turn now to the quantum constraints and their solutions. The first 
step is defining the quantum counterparts of the classical constraints (\ref{vector},\ref{scalar},\ref{gauss}). In LQG we assume, that the quantum Gauss constraints corresponding to the classical expression in (\ref{gauss}) still generate the ``Yang-Mills'' gauge transformations, hence their solutions are functions such that 
\be  
\Psi(\phi,a^{-1}Aa+a^{-1}da)\ =\ \Psi(\phi,A)\label{solgauss} 
\ee
for every $a:M\rightarrow $SU(2).

Similarly, we assume that the quantum vector constraints generate the local diffeomorphism transformations of the quantum states, and in the consequence, the quantum vector constraint carries over to the condition that $\Psi$ be invariant with respect to all local diffeomorphisms  $\varphi: M\rightarrow M$, that is
\be 
\Psi(\varphi^*\phi,\varphi^*A)\ =\ \Psi(\phi,A).\label{solvc}
\ee 

The quantum deparametrized scalar constraint operator has the following form,
\be 
\hat{C}'(x) \Psi\ =\ \left(\hat{\pi}(x)\ - \ \hat{h}(x)\right)\Psi. \label{qDSCo}
\ee  
We use the equation (\ref{hclas}) (which gives the expression for $h$ as a functional of $A$ and $P$) to quantize the second term in the parenthesis. Heuristically we get
\ben 
\hat{h}(x)\ =\ h(\hat{A},\hat{P})(x).
\een
 Due to operator ordering aspects the definition of $\hat{h}$ is not unique and will be completed later in this paper. In order to avoid a quantum anomaly we must respect the classical symmetry in (\ref{commclas}) also at the quantum level and must make sure, that 
\be 
[\hat{h}(x),\hat{h}(y)]=0, \label{comm}
\ee 
(compare to (\ref{commclas})).
Given the quantum constraint operator (\ref{qDSCo}), the constraint itself 
reads
\be 
\left(\hat{\pi}(x)\ - \ \hat{h}(x)\right)\Psi \ =\ 0. \label{qDSCo'}
\ee  
To solve the quantum deparametrized scalar constraint, we write $\Psi$ as 
\be 
\Psi\ =\ e^{i\int d^3x\hat{\phi}(x)\hat{h}(x)}\psi,\label{ehpsi}
\ee
with a new function $\psi$, and insert it in (\ref{qDSCo'}) to obtain
\be 
\frac{\delta}{\delta\phi(x)}\Psi(\phi,A)\ =\ i\hat{h}(x)\Psi(\phi,A).\label{theequation} 
\ee

Due to the commutator in  (\ref{comm}), the constraint equation (\ref{theequation}) turns into 
\be 
\frac{\delta}{\delta\phi(x)}\psi\ =\ 0.  
\ee
Hence, a general solution to (\ref{theequation}) is 
\be \Psi(\phi,A)\ =\ e^{i\int d^3x\hat{\phi}(x)\hat{h}(x)}\psi(A).\label{solsc}
\ee
Notice, that the exponentiated operator acting at $\psi$ on the right hand side of (\ref{ehpsi})  is  Yang-Mills gauge, and diffeomorphism invariant itself. Therefore: 

\bigskip

{\it A general solution to the  quantum vector, gauss and  scalar constraints is every function (\ref{solsc}), such that for every local diffeomorphism $\varphi:M\rightarrow M$,}
\be 
\psi(\varphi^*A)\ =\ \psi(A), 
\ee  
{\it and for every $a:M\rightarrow$SU(2)}
\be 
\psi(a^{-1}Aa+a^{-1}da)\ =\ \psi(A).
\ee
\medskip

In the remaining part of the article we will be using the abbreviation
\be  
\int d^3x\hat{\phi}\hat{h} \ :=\ \int d^3x\hat{\phi}(x)\hat{h}(x).
\ee

\subsection{Quantum Dirac observables} 
A quantum Dirac observable is the restriction to the space of solutions to the quantum constraints of an operator ${\cal O}$ which satisfies the following two properties:
\begin{itemize}
\item $\hat{\cal O}$ is  invariant under local diffeomorphism and Yang-Mills gauge transformations, 
\item 
\be [\hat{\cal O},\hat{C}'(x)]\ =\ 0
\label{[O,C']}\, .\ee
\end{itemize}
Following the ideas of the relational framework for observables \cite{Rov1,Ditt1,Ditt2}
it is easy to construct a large family of Dirac observables. Let $\hat{L}$ be a linear operator  which maps  the functions $A\mapsto \psi(A)$  into  functions $A\mapsto \hat{L}\psi(A)$. Consider an operator 
\be 
{\cal O}(\hat{L})\ :=\ e^{i\int d^3x\hat{\phi}\hat{h}}\hat{L}e^{-i\int d^3x\hat{\phi}\hat{h}}.\label{OL} 
\ee
As required, the operator ${\cal O}(\hat{L})$ commutes with the quantum version of the deparametrized scalar constraints,
\be 
[{\cal O}(\hat{L}),\hat{C}'(x)]\ =\ 0.
\ee
Moreover, the operator ${\cal O}(\hat{L})$ is Yang-Mills gauge and local diffeomorphism invariant provided the operator $\hat{L}$ is.  

Each of the operators ${\cal O}(\hat{L})$ defined by a Yang-Mills gauge, and diffeomorphism invariant operator $\hat{L}$ preserves the space of solutions to the constraints. Indeed,
\ba \label{scobssol}
{\cal O}(\hat{L})e^{i\int d^3x\hat{\phi}\hat{h}}\psi(A)  \ &=&\ e^{i\int d^3x\hat{\phi}\hat{h}}\psi'(A),\\
\psi'\ &=&\ \hat{L}\psi.\nonumber
\ea
The operators (\ref{scobssol}) with the Yang-Mills gauge, and local diffeomorphism invariant operators $\hat{L}$ set a family (algebra, modulo the domains) of the Dirac observables. 
The total scalar field momentum $\int_Md^3x \hat{\pi}(x)$  also  defines one of the quantum Dirac observables (\ref{scobssol}), namely
\be 
{\cal O}(\int_Md^3x \hat{h}(x))\ =\ \int_Md^3x \hat{h}(x). 
\ee

The family of the Dirac observables (\ref{scobssol})  in fact contains {\it all} the quantum Dirac observables. To see that this is true, suppose an operator $\hat{{\cal O}}$ satisfies the condition (\ref{[O,C']}) at each $x\in M$. 
Let us write the operator in the following way 
\be 
\hat{{\cal O}}\ =\ e^{i\int d^3x\hat{\phi}\hat{h}}\hat{K}e^{-i\int d^3x\hat{\phi}\hat{h}},\label{OM} 
\ee
where $\hat{K}$ is a priori arbitrary operator. The condition (\ref{[O,C']})
with $\hat{{\cal O}}$ substituted for the right hand side of (\ref{OM}), takes the following form 
\be 
[\hat{K},\hat{\pi}(x)]\ =\ 0.\label{[M,pi]}
\ee
The set of all the solutions $\hat{K}$ to (\ref{[M,pi]}) is generated by the following solutions: $(i)$ given any $x\in M$,   
$$ \hat{K}=\hat{\pi}(x),$$ 
and $(ii)$  
$$\hat{K}=\hat{L},$$ 
considered above, that is $\hat{L}$ which  maps  the functions $A\mapsto \psi(A)$  into  functions $A\mapsto \hat{L}\psi(A)$. 

The solutions of the type $(ii)$  give rise exactly to the family of the 
quantum Dirac observables (\ref{scobssol}) we have introduced above.    
On the other hand, a solution of the type $(i)$ gives rise to the following
quantum   Dirac observable
\be e^{i\int d^3x\hat{\phi}\hat{h}}\hat{\pi}(x)e^{-i\int d^3x\hat{\phi}\hat{h}}\ =\
\hat{\pi}(x)\ -\ \hat{h}(x). \label{Opi} \ee 
However, we  should keep in mind that what really defines  a quantum Dirac observable is the restriction to the space of solutions to the quantum constraints.  The restriction of (\ref{Opi}) is identically zero. This shows that all the quantum Dirac observables are those defined by (\ref{scobssol}) and diffeomorphism and Young-Mills invariant operator $\hat{L}$.

\subsection{Classical interpretation of the Dirac observables}
Suppose, that a given operator $\hat{L}$ used to construct the Dirac observable ${\cal O}(\hat{L})$ corresponds in the quantum theory to a classical function $L$ defined on the gravitational phase space $\Gamma_{\rm gr}$, and that the support of $L$ is contained in the set on which 
\be 
\Cgr\ <\ 0. 
\ee
To find a classical observable ${\cal O}(L)$ whose quantum counterpart is 
${\cal O}(\hat{L})$, it is convenient to express the operator (\ref{OL}) in terms of a formal power series given by
\be
{\cal O}(\hat{L})\ = \sum\limits_{n=0}^{\infty} \frac{i^n}{n!}
\left[\hat{L}, \int d^3x\hat{\phi}\hat{h}\right]_{(n)}
\ee
where $[.,.]_{(n)}$ denotes the iterated commutator defined by $[\hat{L}, \int d^3x\hat{\phi}\hat{h}]_{(0)}=\hat{L}$ and $[\hat{L}, \int d^3x\hat{\phi}\hat{h}]_{(n)}=[[\hat{L},\int d^3x\hat{\phi}\hat{h}]_{(n-1)},\int d^3x\hat{\phi}\hat{h}]$.
The usual  substitution $[\cdot,\cdot]\mapsto -i\{\cdot,\cdot\}$, leads to a formal  series
\be
{\cal O}(L) = \sum\limits_{n=0}^{\infty}\frac{1}{n!}\{L,\int d^3x\phi h\}_{(n)}\label{OL}
\ee
for a classical observable  ${\cal O}(L)$.  That series is very well known in the theory of  relational observables \cite{Ditt1,Ditt2,ThiemannRed,GieThiemm}. 
To recall its meaning we  first consider a  slightly more general expression with $\phi$ replaced by a point dependent parameter $M\ni x\mapsto t(x)$, namely
\be \alpha_t^*(L)\ =\ \sum\limits_{n=0}^{\infty}\frac{1}{n!}\{L,\int d^3x t h\}_{(n)}.\ee 
The $*$ denotes the pullback, and the map 
$$\alpha_t:\Gamma_{\rm gr}\rightarrow\Gamma_{\rm gr}$$
is defined by the hamiltonian flow  $\beta_t:\Gamma\rightarrow \Gamma$ generated 
in the full phase space $\Gamma$ by the constraints $C'(x)$  with the parameters $t(x)$. The action of the flow reads,
$$\beta_t(A,P,\phi,\pi)\ =\ (\alpha_t(A,P),\phi-t,\pi). $$
Clearly 
\be \beta_\phi(A,P,\phi,\pi)\ =\ (\alpha_\phi(A,P),0,\pi). \label{betaphi}\ee 
The value of ${\cal O}({L})$ at any point $(A,P,\phi,\pi)$ is defined to be
\be
{\cal O}({L})(A,P,\phi,\pi)\ =\  {L(\alpha_\phi(A,P))}.
\ee

In conclusion, the quantum Dirac observable ${\cal O}(\hat{L})$  corresponds to the classical function,  that is also an observable,  ${\cal O}({L})$,
\be \widehat{{\cal O}(L)}\ =\ {\cal O}(\hat{L}).\ee

At this point a comment  about the status of the operator ${\cal O}(\hat{L})$
is  appropriate. It may happen, that given a point $(A,P,\phi,\pi)$ in the classical phase space,  the series (\ref{OL}) is non-converging.   In fact we encounter cases like that in the LQC models of homogeneous isotropic universe with positive
cosmological constant \cite{KP10}. However, still the operator 
${\cal O}(\hat{L})$ is well defined as long as a self adjoint extension
for the operator $\int d^3x\phi\hat{h}$ is fixed, and therefore the unitary operator 
${\rm exp}(i\int d^x\phi\hat{h})$ is well defined. Then, the quantum evolution just goes beyond
the classical theory. That is exactly the reason, why we have chosen 
to define the  Dirac observables directly in the quantum theory, and only interpret them in the classical theory as secondary objects.

\subsection{Dynamical evolution of the observables}

 The Dirac observables we have defined are relational observables (often called ``partial'' \cite{Rov1},\cite{Ditt1}, section I.2 of \cite{Thiemm2} ).
 For that class of observables one is able to define a non -- vanishing evolution generated by a so called physical Hamiltonian, that will be introduced in the next section.
The dynamics is defined with respect to an internal time given by the values, which that field $\phi$ takes while being transformed along its gauge orbit. This can be seen in the construction of the quantity
${\cal O}({L})$ from a given function $L$ by generalizing the choice of the evaluation point from (\ref{betaphi}) to 
\be 
\beta_{\phi-\phi_0}(A,P,\phi,\pi)\ =\ (\alpha_{\phi-\phi_0}(A,P),\phi_0,\pi), 
\ee 
where $\phi_0$ is an arbitrarily fixed function on $M$. We denote the resulting function defined on the phase space $\Gamma$ by ${\cal O}_{\phi_0}(L)$, that is
\be {\cal O}_{\phi_0}(L)(A,P,\phi,\pi)\ =\ L(\alpha_{\phi-\phi_0}(A,P)). \ee
For the function  ${\cal O}_{\phi_0}(L)$ to be well defined, the 
 flow $\beta_t:\Gamma\rightarrow\Gamma$ has to be well defined for  
$$t=\phi-\phi_0$$ 
in the domain of the function $L$. 

That classical construction leads to a corresponding
quantum operator definition
\be \label{Opphi}
{\cal O}_{\phi_0}(\hat{L})\Psi(\phi,A)\ =\ e^{i\int d^3x(\phi(x)-\phi_0(x))\hat{h}(x)}\hat{L} e^{-i\int d^3x(\phi(x)-\phi_0)\hat{h}(x)}\Psi(\phi,A),
\ee
where we used $\hat{\phi}\Psi(\phi,A)=\phi\Psi(\phi,A)$.
This definition will  not enlarge the class of the Dirac observables (\ref{scobssol}), indeed
\be 
{\cal O}_{\phi_0}(\hat{L})\ =\ {\cal O}(\hat{L}') \label{scdyn1}
\ee
with
\be 
\hat{L}'\ =\ e^{-i\int d^3x\phi_0(x)\hat{h}(x)}\hat{L} e^{i\int d^3x\phi_0\hat{h}(x)}. \label{scdyn2} 
\ee
In this way, in the  algebra of the (formal) solutions to the condition
\be [\hat{O},\hat{C}'(x)]\ =\ 0 \ee
we have defined  an abelian group of automorphisms labelled by the functions $\phi_0$ defined on $M$, namely
\be 
{\cal O}(\hat{L})\ \mapsto\ {\cal O}_{\phi_0}(\hat{L}).\label{evol} 
\ee
If we want to restrict the automorphisms to the algebra of the quantum Dirac
observables, we encounter an obstacle. Given a function $\phi_0$,
we want the operator (\ref{scdyn2}) to be diffeomorphism invariant for every
diffeomorphism invariant operator $\hat{L}$. For the operators $\hat{h}(x)$
that will be constructed from the LQG framework, that condition can be satisfied
only for a constant function,   
\be 
\phi_0(x) = \phi_0\in \mathbb{R}, {\rm for\ \ every}\ \ x\in M\,.
\ee
The result is a 1-dimensional group of automorphisms of the algebra of the quantum Dirac observables.  The group encodes the dependence on the internal time of the algebra of the quantum Dirac observables.

\subsection{The physical Hamiltonian}

The dynamics is generated by the following equation   
\be 
\frac{d}{d\phi_0}{\cal O}_{\phi_0}(\hat{L})\ =\ -i[\hat{h}_{\rm phys},{\cal O}_{\phi_0}(\hat{L})]\label{dynam} 
\ee
Where
\be 
\hat{h}_{\rm phys}\ :=\ \int d^3x \hat{h}(x)\ 
\ee
is usually called the physical hamiltonian for the reason that it is a non-vanishing Dirac observable generating true 'physical' evolution in contrast to the Hamiltonian  constraint.

The physical Hamiltonian will be an exact implementation of the heuristic formula   
\be 
\hat{h}_{\rm phys}\ =\ \int d^3x\sqrt{-\sqrt{\hat{q}}{\hat{C}}^{gr}+\sqrt{\hat{q}}\sqrt{({\hat{C}}^{gr})^2 -\hat{q}^{ab}{\hat{C}}^{gr}_a{\hat{C}}^{gr}_b}}. 
\ee 

We remember however, that the operator will be applied to diffeomorphism invariant states (\ref{solvc}) whereas the operator $\hat{C}^{gr}_a$ should generate the diffeomorphisms. Therefore, assuming the suitable choice of the ordering, the physical Hamiltonian acting on the diffeomorphism
invariant functions $\psi$ is

\be 
\hat{h}_{\rm phys}\,\psi(A)\ =\ \int d^3x\sqrt{-2\sqrt{\hat{q}}\ {\hat{C}}^{gr}}\,\psi(A), 
\ee 
where we also took into account (recall  (\ref{Cgrclass<0})), 
\be  {\hat{C}}^{gr}\ <\ 0.\label{Cgr<0}  \ee
This result coincides with that of \cite{RS}.

\subsection{The Hilbert product between the solutions: $\Hphy$}

Suppose we have a sesquilinear scalar product for the Yang-Mills gauge and local diffeomorphism invariant functions (or distributions) defined on the space of the Ashtekar-Barbero connections. Denote the product of the functions $\psi$ and $\psi'$ by
\be
(\psi|\psi'), 
\ee 
and the corresponding Hilbert space by $\Hil_{\rm diff}$.  

We can use it to define the ``physical'' (that is respecting the dynamics) Hilbert product in the space of solutions (\ref{solsc}):
\be 
\left(e^{i\int \hat{\phi}\hat{h}}\psi\,|\,e^{i\int \hat{\phi}\hat{h}}\psi'\right)_{\rm phys}\ :=\ (\psi|\psi'). 
\ee  

The resulting Hilbert space $\Hphy$ is  ``physical'', and its elements are the physical states.  

\subsection{Summary:  the exact structures we need}

In summary, in order to construct the quantum model we will need:

\begin{itemize}
\item the Hilbert space $\Hil_{\rm diff}$ of the Yang-Mills gauge and the local diffeomorphism invariant quantum states of geometry, 
\item the operators in $\Hil_{\rm diff}$ which admit a well understood geometric interpretation,
\item the physical Hamiltonian operator $\hat{h}_{\rm phys}$ defined in a suitable domain in $\Hil_{\rm diff}$ (which is not expected to be dense, because the heuristic formula for the operator involves the square roots of non definite expressions).
\end{itemize}
              
Given all that, the physical Hilbert space is unitarily isomorphic via 
\be 
e^{i\int d^3x\hat{\phi}\hat{h}}\psi\ \mapsto\ \psi\label{physkin}
\ee
with the domain of $\hat{h}_{\rm phys}$ in $\Hil_{\rm diff}$.

Every observable ${\cal O}(\hat{L})$ (for simplicity let $\hat{L}$ be bounded) is the pullback by (\ref{physkin}) of an operator $\hat{L}$ which preserves the completion of the domain of $\hat{h}_{\rm phys}$.

Finally,  the emerged dynamical evolution (\ref{dynam}) of the observables reads
\be 
 \hat{L}(\tau)\ =\ e^{-i\tau\hat{h}_{\rm phys}}\hat{L}e^{i\tau\hat{h}_{\rm phys}}.
\ee
This is precisely the very well known Heisenberg picture evolution defined by the Hamiltonian 
$\hat{h}_{\rm phys}$.

Notice, that in fact, it is not necessary for $\hat{L}$ to preserve the domain of $\hat{h}_{\rm phys}$. Indeed, given any $\psi$ in that domain, the expectation value 
\ben
(\psi|e^{-i\tau\hat{h}_{\rm phys}}\hat{L}e^{i\tau\hat{h}_{\rm phys}}\psi)\ =\ (e^{i\tau\hat{h}_{\rm phys}}\psi\,|\,\hat{L}e^{i\tau\hat{h}_{\rm phys}}\psi)
\een
is well defined. This can be seen by using that it  is equivalent to replace $\hat{L}$ by the operator 
\be 
\hat{L}'\ =\  P\hat{L}P, 
\ee  
where $P$ is the orthogonal projection onto the completion of the domain of $\hat{h}_{\rm phys}$, and to considering the pullback of the Dirac observable ${\cal O}(\hat{L}')$ together with its dynamics.           

\medskip

This kind of structure  will be necessary for the outcome. This is all we need to complete the quantization of  a model of quantum gravity coupled to a scalar field.

In the derivation of the operator corresponding to the physical hamiltonian $\hat{h}_{\rm phys}$, however, we will need yet more structure:
\begin{itemize}
\item the operator $\hat{h}_{\rm phys}$ should be defined by using the suitably defined  operator valued distribution $M\ni x\ \mapsto\ \widehat{\sqrt{q}(x)\Cgr(x)}$, 
\item the distribution should be self-adjoint, so that we can use the spectral decomposition to define the subspace 
\be 
\widehat{\sqrt{q}(x)\Cgr(x)}\ <\ 0 
\ee 
and thereon the new operator valued distribution 
\be 
\hat{h}(x)\ =\ \sqrt{-2\sqrt{q}(x)\Cgr(x)}, 
\ee 
\item we should be able to verify the condition
\be 
[\hat{h}(x),\hat{h}(y)]\ =\ 0,\label{[h,h]}
\ee
\item and finally define
\be 
\hat{h}_{\rm phys}\ =\ \int d^3x \hat{h}(x). 
\ee  
\end{itemize}

Notice, that none of the operators $\sqrt{q}(x)\Cgr(x)$ or $\hat{h}(x)$ can be defined within the Hilbert space $\Hil_{\rm diff}$, because the $x$ dependence manifestly breaks the diffeomorphism invariance. Therefore, the properties of the self-adjointness require some extra Hilbert spaces, $\Hil_{{\rm diff},x}$, labelled by the points of $M$, whereas  the commuting at different points can be defined only on a yet bigger Hilbert space.    

Remarkably, all the suitable structures can be constructed within the LQG framework, as we will explain in the next section.

\section{Application of LQG} 

\subsection{The Hilbert spaces}

\subsubsection{The kinematical Hilbert space of quantum states of the geometry} 

In LQG (we use the notation of  \cite{AshLew1}), the kinematical Hilbert space of quantum states of the geometry is set by the so called cylindrical functions of the connection $A$. A cylindrical function  is defined by a set $\alpha$ of finite curves  $e_1,...,e_n$ in $M$ and by a continues function $f:{\rm SU}(2)^n\rightarrow \mathbb{C}$, in the following way 
\be 
\psi_{\alpha,f}(A)\ =\ f(A(e_1),...,A(e_n))\label{cyl} 
\ee
where the symbol $A(e)$ denotes the parallel transport along $e$ defined by the connection $A$. The set Cyl of the cylindrical functions is a vector space, and an associative algebra. The space of the cylindrical functions ${\rm Cyl}$ is endowed with an integral 
\be
\psi_{\alpha,f}\mapsto \int\psi_{\alpha,f} 
\ee 
used to define the sesquilinear scalar product  
\be 
(\psi_{\alpha,f}|\psi_{\alpha',f'})_{\rm gr}\ =\ \int \overline{\psi_{\alpha,f}}\psi_{\alpha',f'}, 
\ee 
and defines (after the completion) the kinematical Hilbert space $\Hil$ for the geometric operators in LQG. We assume in this paper that the manifold and the curves are piecewise analytic. Then, for every cylindrical function there exist curves $\alpha =\{e_1,...,e_n\}$ which form a graph embedded in $M$ (that is they are allowed to intersect only at the ends) such that the function is given by (\ref{cyl}). The curves $e_I$ are called edges of the given graph $\alpha$.\footnote{To be more precise, in what follows, an edge is either an oriented semianalytic imbedding of a circle in $M$, or a parametrization free, oriented, finite curve defined by $ e:[0,1]\rightarrow M $ such that either $e$ is  an imbedding, or $e(0)=e(1)$.} 

For a cylindrical function defined by a graph, we have
\be 
\int\psi_{\alpha,f}\ =\ \int_{{\rm SU(2)}^n} d^ngf(g_1,...,g_n),
\ee            
where $d^ng$ is the Haar measure on SU(2)$^n$. The geometric operators preserving the space Cyl are
\be 
\hat{A}(e)^B_C\psi_{\alpha,f}(A)\ =\ {A}(e)^B_Cf(A(e_1),...,A(e_n))
\ee
and
\be 
\int_S \hat{P}^a_i\psi_{\alpha,f}\ =\ \frac{1}{2i}\int_S \frac{\delta}{\delta A^i_a}\psi_{\alpha,f}\ \eta_{abc}dx^b\wedge dx^c. 
\ee

There is an orthogonal decomposition of $\Hil$ into subspaces $\Hil'_\alpha$ labelled by the embedded graphs $\alpha$. To define it, denote first by (unprimed) $\Hil_\alpha\subset\Hil$ the Hilbert subspace spanned by  the cylindrical functions $\psi_{\alpha,f}$, with all the possible functions $f$. Those spaces, however, are too big to provide the orthogonal decomposition. Given a graph $\alpha$, whenever a graph $\beta$ can be obtained from  the edges of $\alpha$ by glueing, or reversing the orientation or removing some of them, then $\Hil_\beta\subset\Hil_\alpha$. Therefore, define $\Hil'_\alpha\subset \Hil_\alpha$ to be the orthogonal complement in $\Hil_\alpha$ of the subspace spanned by those
subspaces $\Hil_\beta$. The decomposition is
\be 
\Hil\ =\ \bigoplus_{\alpha}\Hil'_{\alpha},\label{decomp}
\ee     
where $\alpha$ runs through the set of all the semianalytic embedded graphs in $M$.

\subsubsection{The Hilbert space of the diffeomorphism invariant states of the geometry} 

Semianalytic diffeomorphisms Diff$(M)$ of  $M$ preserve the space Cyl and act unitarily in the Hilbert space $\Hil$ just by the natural pullback of the Ashtekar-Barbero connections. Denote the action of $\varphi\in$Diff$(M)$ by
\be 
U_\varphi:\Hil\rightarrow\Hil.
\ee    
To implement the construction of the quantum operator corresponding to the physical Hamiltonian, we will need two different Hilbert spaces: One of them includes states, that are invariant with respect to all (semi-analytic) local diffeomorphisms Diff$(M)$ of $M$ and the other one is the home of the states invariant with respect to the subgroup Diff$(M,x)$, which  preserves a given point $x\in M$. (Later, we will also impose the Gauss constraint, that is the condition of Yang-Mills gauge invariance). Let Diff stands for either Diff$(M)$ or  Diff$(M,x)$. The only Diff invariant direction in $\Hil$ is the constant function. However, since the group Diff is not compact, we expect the  invariant states to be  distributions on the space of the Ashtekar-Barbero connections, that is linear maps
\be 
\langle \Psi|:{\rm Cyl}\rightarrow \mathbb{C}.
\ee
Whereas the space of all distributions seems to be too big, a suitable rigging map can be defined, which carries each $\psi\in {\rm Cyl}$ into a Diff invariant distribution $\eta_{\rm Diff}(\psi)$. To recall the definition of this map, we need the orthogonal decomposition (\ref{decomp}). The map $\eta_{\rm Diff}$  is  introduced  for each subspace $\Hil'_\alpha$ individually. By the linearity, it extends to
every cylindrical function. That is, the  domain of the rigging map $\eta_{\rm Diff}$ is Cyl$\subset \Hil$. The first step in the construction of the rigging map $\eta_{\rm Diff}$, is identification of the elements of  $\Hil'_\alpha$ that will be annihilated. Consider those diffeomorphisms $\varphi\in$Diff which map each edge of $\alpha$ into another edge modulo the orientation, and let us call them the symmetries of $\alpha$ and denote their group by Diff$_{\alpha}$. The functions $\psi\in \Hil'_\alpha$ invariant with respect to Diff$_{\alpha}$ form a subspace denoted either by $\Hil'_{\alpha,{\rm inv}}$ in the Diff$=$Diff$(M)$ case, or  $\Hil'_{\alpha,{\rm inv},x}$, in the case of Diff$=$Diff$(M,x)$. The elements of  $\Hil'_\alpha$ orthogonal to $\Hil'_{\alpha,{\rm inv}}$ are  annihilated by the rigging map $\eta_{\rm Diff}$. For $\psi\in \Hil'_{\alpha,{\rm inv}}$, $\eta_{\rm Diff}(\psi)$ is defined as follows
\be 
\eta_{\rm Diff}(\psi): \psi''\mapsto \sum_{[\phi]\in {\rm Diff}/{\rm Diff}_{\alpha}}(U_\phi\psi|\psi'').\label{eta}
\ee

Note, that if $\psi''\in\Hil'_{\alpha''}$, then the right hand side is zero if $\alpha''$ is not Diff equivalent to $\alpha$, and in the case there is $\phi''\in$Diff such that 
\be 
\phi''(\alpha)=\alpha'', 
\ee  
the only possibly non-zero term in the sum in (\ref{eta}) is
\be 
\eta_{\rm Diff}(\psi):\psi''\mapsto (U_{\phi''}\psi|\psi''). 
\ee
     
Since every cylindrical function is a finite sum of elements of the Hilbert spaces $\Hil'_\alpha$, $\eta_{\rm Diff}(\psi)$ is defined in Cyl. For the same reason, the map
\be 
\psi\mapsto \eta_{\rm Diff}(\psi)
\ee 
extends by the finite linearity to Cyl. 

With the rigging map $\eta_{\rm Diff}$ we define not only the vector space of the Diff invariant states to be the image $\eta_{\rm Diff}({\rm Cyl})$, but also the sesquilinear product
\be
(\eta_{\rm Diff}(\psi)|\eta_{\rm Diff}(\psi'))_{\rm Diff}\ :=\ \langle\eta_{\rm Diff}(\psi),\psi'\rangle 
\ee

\medskip

In this way we have defined a Hilbert space $\Hil_{\rm Diff}$. The map $\eta_{\rm Diff}$ defines a natural isometry 
\be 
\Hil_{\rm Diff}\ \equiv\ \bigoplus_{[\alpha]}\Hil'_{\alpha,{\rm Diff}}
\ee
where $[\alpha]$ runs through the set of the Diff classes of the graphs embedded in $M$. Recall that Diff $=$ Diff$(M)$, Diff$(M,x)$. Therefore, we have defined 
two types of the Hilbert spaces: the Hilbert $\Hil_{{\rm Diff}(M)}$ and, respectively, per each point $x\in M$, the Hilbert space $\Hil_{{\rm Diff}(M,x)}$.

\subsubsection{The Hilbert spaces of the Yang-Mills gauge and diffeomorphism invariant states of the geometry}

Imposing the Gauss constraint is yet easier, than requiring diffeomorphism invariance, and could be equivalently done, either before, or after solving the diffeomorphism constraint. The group of unitary transformations of $\Hil$ given by the Yang-Mills gauge transformations is compact. Hence all solutions to the Gauss constraint in $\Hil$ are invariant elements of $\Hil$ (as opposed to non-normalizable states, distributions). Moreover, the group of the Yang-Mills gauge transformations (\ref{solgauss}) preserves each of the subspaces $\Hil'_{\alpha}$. For every Yang-Mills gauge invariant $\psi\in$Cyl, the Diff invariant distribution 
\be \eta_{\rm Diff}(\psi)\in \Hil_{\rm Diff} \ee     
is also insensitive to gauge transformations of $\psi''\in$Cyl. Namely, 
the number
\ben
\eta_{\rm Diff}(\psi)(\psi'')
\een
is invariant. The converse is also true: If $\eta_{\rm Diff}(\psi)(\psi'')$ is invariant with respect to the Yang-Mills gauge transformations of $\psi''$, than $\psi$ is Yang-Mills gauge invariant. 

\medskip

In conclusion, the Yang-Mills gauge and diffeomorphism invariant distributions on the space of the Ashtekar-Barbero connections we want to use to construct the Hilbert space $\Hil_{\rm diff}$ of section III.G, are the distributions 
\be 
\eta_{{\rm Diff}(M)}(\psi) 
\ee
obtained from the Yang-Mills gauge invariant cylindrical functions $\psi$. Denote their Hilbert space by $\Hil_{\rm diff}$. By construction
\be 
\Hil_{\rm diff}\subset \Hil_{{\rm Diff}(M)}.
\ee
 
For the introduction of the physical Hamiltonian we will also use the Hilbert space $\Hil_{{\rm diff},x}$ obtained by replacing in the construction of the Hilbert space $\Hil_{\rm diff}$ the group Diff$(M)$ by the group Diff$(M,x)$.

\subsection{The operators}

\subsubsection{The Dirac observables}

From the previous subsection we already have the LQG candidate for the Hilbert space  $\Hil_{\rm diff}$ of the Yang-Mills gauge invariant and diffeomorphism invariant quantum states of geometry. As we already know from  Section III.G, from a suitable subspace of this space we will construct the ``physical'' Hilbert space of solutions to  all the constraints of the model we are considering. Secondly, in the Hilbert space $\Hil_{\rm diff}$ we will need the operators representing the geometry of the initial data defined on $M$, from which we will construct the Dirac observables. 

Let us begin with this second task, because it is easier. We assume below, that the operators we consider in the Hilbert space $\Hil$, as the domain have the vector subspace Cyl of the cylindrical functions. Every Yang-Mills gauge and Diff$(M)$ symmetric  operator $\tilde{L}$ defined in the kinematical Hilbert space $\Hil$, defines naturally by the duality a symmetric operator $\hat{L}$ in $\Hil_{\rm diff}$,
\be 
\langle\hat{L}\eta_{{\rm Diff}(M)}(\psi),\psi''\rangle\ :=\ \langle\eta_{{\rm Diff}(M)}(\psi),\tilde{L}\psi''\rangle\ =  \langle\eta_{{\rm Diff}(M)}(\tilde{L}\psi),\psi''\rangle 
\ee
where the bracket is the action of a distribution (a first entry) into a given cylindrical function $\psi''$, that is, we could  phrase it in a simpler way
\be
\hat{L}\eta_{{\rm Diff}(M)}(\psi)\ =\ \eta_{{\rm Diff}(M)}(\hat{L}\psi).\label{Leta}
\ee
An excellent example of a Yang-Mills gauge and diffeomorphism invariant operator in $\Hil$ available in the literature \cite{AshLew1},\cite{AshLew2} is the volume of the underlying manifold $M$ operator 
\be 
\tilde{V}_M\ =\ \int d^x \tilde{\sqrt{q}}(x). 
\ee 
Another example we manage to construct might be any  quantum operator  representing the integral of a scalar constructed from the intrinsic or extrinsic curvature.

In the kinematical Hilbert space $\Hil$, there is also a well defined operator valued distribution 
\be 
\tilde{\sqrt{q}}(x)\ =\ \sum_{x'\in M}\delta(x,x')\tilde{\sqrt{q}}_{x'},
\ee
where each of the operators $\tilde{\sqrt{q}}_{x'}$ is Diff$(M,x')$ invariant. The uncountable sum on the right hand side is well defined, because for every smearing function $F:M\rightarrow \mathbb{R}$, and a cylindrical function $\psi_{\alpha,f}$, we have
\be 
\int d^3x F(x)\tilde{\sqrt{q}}(x)\,\psi_{\alpha,f}\ =\ \sum_{I=1}^nF(v_I)\tilde{\sqrt{q}}_{v_I}\psi_{\alpha,f},\label{sumv}
\ee
where $v_1,...,v_n$ are the vertices of the graph $\alpha$. Via (\ref{Leta}), for every $x'\in M$, the operator $\tilde{\sqrt{q}}_{x'}$ defines an operator $\widehat{\sqrt{q}}_{x'}$ in $\Hil_{{\rm diff},x'}$. Morally, $\tilde{\sqrt{q}}(x)$  is also Diff$(M,x)$ invariant for every given $x\in M$, therefore  (\ref{Leta}) should also be somehow generalized to this case. Indeed,  (see \cite{MarLew}) the suitable generalization is natural and provides in this case a distribution 
\be  
\widehat{\sqrt{q}}(x)\ =\ \sum_{x'\in M}\delta(x,x')\widehat{\sqrt{q}}_{x'},
\ee
which makes sense due to the fact that all the Hilbert spaces $\Hil_{{\rm diff},x}$ are embedded in the single vector space Cyl$^*$. 

There is one more technical remark in order at this point. Consider two operator valued distributions  in $\Hil$, of the form
\be 
\tilde{A}(x)\ =\ \sum_{x'\in M}\delta(x,x')\tilde{A}_{x'}, \ \ \ \tilde{B}(x)\ =\ \sum_{x'\in M}\delta(x,x')\tilde{B}_{x'}  
\ee   
each of which satisfies the property (\ref{sumv}). A natural regularization by smearing leads to a new operator valued distribution
\be 
\sqrt{\tilde{A}(x)\tilde{B}(x)}\ =\ \sum_{x'\in M}\delta(x,x') \sqrt{S(\tilde{A}_{x'}\tilde{B}_{x'})}
\ee  
which also has the property (\ref{sumv}), where $S$ stands for a symmetric product of the operators, and the domain of the resulting operator is restricted by the positivity of $S(\tilde{A}_x\tilde{B}_x)$ requirement. The regularization consist in the smearing 
\be 
\tilde{A}_\epsilon(x)\ =\ \int d^3 \tilde{A}(y) \delta_{\epsilon}(y,x), \ \ \ \tilde{B}_\epsilon(x)\ =\ \int d^3 \tilde{B}(y) \delta_{\epsilon}(y,x)
\ee
with a smearing function whose support goes uniformly to $x=y$ as $\epsilon\rightarrow 0$, and which goes  to  the Dirac $\delta(x,y)$. The key trick is an observation that for every fixed graph $\alpha$, for sufficiently small $\epsilon$  
\be 
\tilde{A}_\epsilon(x)\tilde{B}_\epsilon(x)\psi_{\alpha,f}\ =\ \sum_{I=1}^n (\delta_\epsilon(x,v_I))^2\tilde{A}_{v_I}\tilde{B}_{v_I}\psi_{\alpha,f}, 
\ee
for any cylindrical function $\psi_{\alpha,f}$, and more over, the sum on the right hand side contains at most one non zero element. Due to the latter property        
\be 
\sqrt{\tilde{A}_\epsilon(x)\tilde{B}_\epsilon(x)}\ =\ \sum_{I=1}^n \delta_\epsilon(x,v_I)\sqrt{\tilde{A}_{v_I}\tilde{B}_{v_I}}\psi_{\alpha,f} 
\ee
provided the square root is well defined itself. Finally,
\be 
\int d^3x F(x)\sqrt{\tilde{A}_\epsilon(x)\tilde{B}_\epsilon(x)}\psi_{\alpha,f}\ \rightarrow\ \sum_{I=1}^nF(v_I)\sqrt{\tilde{A}_{v_I}\tilde{B}_{v_I}}\psi_{\alpha,f}. 
\ee 
  
\subsubsection{The quantum scalar constraint and the physical Hamiltonian}  

As we remember, our first task we can finally turn to now, is a construction of the physical Hamiltonian operator 
\be 
\hat{h}_{\rm phys}\ =\ \int d^3x \sqrt{-2\widehat{\sqrt{q}(x)C^{\rm gr}}(x)} 
\ee
defined in $\Hil_{\rm diff}$. 

A quantum scalar constraint $\hat{C}_{\rm gr}$ was defined in \cite{Thiemm1}, its properties and possible generalizations  were studied in \cite{MarLew,AshLew1}. We will be using here the formulation of the scalar constraint of \cite{AshLew1}. In order to use it for our current construction, we will need a new element. Thus far, the scalar constraint was used either as smeared against arbitrary laps function $\int d^3xN(x)\hat{\cal C}(x)$, or, as the master constraint $\int d^3x \widehat{\sqrt{q}}(x)^{-1}\hat{C}(x)\hat{C}^\dagger(x)$, or as a physical Hamiltonian defined after deparametrization with respect to 4 scalar fields. The smeared scalar constraint maps a domain in $\Hil_{\rm diff}$ into Cyl$^*$,  there is no sense in which it could be symmetric or self-adjoint. The master constraint, on the other hand,  as well as the physical Hamiltonian after the 4-fold deparametrization, respectively, is defined in the kinematical Hilbert space $\Hil$ as a graph preserving operator. The current case, is a new one, we will need an  operator
$\sqrt{-2\widehat{\sqrt{q}(x)C^{\rm gr}}(x)}$ defined in $\Hil_{\rm diff}$.
     
The quantum scalar constraint presented in \cite{AshLew1} takes the following form,
\be 
\int d^3xN(x)\hat{\cal C}(x)\ =\ \sum_{x\in M}\hat{\cal C}_x, 
\ee  
where each of the operators  $\hat{\cal C}_x$ maps its domain contained in $\Hil_{\rm diff}$ into  $\Hil_{{\rm diff},x}$.  However, as it follows from \cite{MarLew}, it naturally defines an operator in the corresponding Hilbert space $\Hil_{{\rm diff},x}$. The advantage is, that only now we can require the symmetry (self-adjointness) of those operators. As defined in \cite{AshLew1},  the operators $\hat{\cal C}_x$ come out  non-symmetric. The minor improvement, but necessary for our current work, consists in replacing them by symmetric operators 
\be 
\hat{C}^{\rm gr}_x\ =\ \frac{1}{2}\left(\hat{\cal C}_x\ + \hat{\cal C}_x^\dagger\right). 
\ee        
and  choosing an essentially self-adjoint extension that may be non-unique. Then, the quantum scalar constraint operator we will use for the physical Hamiltonian  takes the following form
\be  
\hat{C}^{\rm gr}(x)\ =\ \sum_{x'\in M}\delta(x,x')\hat{C}^{\rm gr}_{x'}.
\ee 
On the other hand we have already considered above the volume density quantum operator written in the similar form, 
\be 
\widehat{\sqrt{\hat{q}}(x)}\ =\ \sum_{x'\in M}\delta(x,x')\widehat{\sqrt{\hat{q}}}_{x'}. 
\ee

At this point, we are in the position to define the operator
\be 
\sqrt{-2\widehat{\sqrt{q}(x){C}^{\rm gr}(x)}}
\ee
A regularization in $\Hil$ similar to the one discussed above, gives (modulo the symmetrization of the product of the operators $\widehat{\sqrt{q_{x'}}}$ and $\hat{C}^{\rm gr}_{x'}$)
\be 
\sqrt{-2\widehat{\sqrt{q}(x){C}^{\rm gr}(x)}}\ =\ \sum_{x'\in M}\delta(x,x') \sqrt{-2\widehat{\sqrt{q_{x'}}}^{1/2}\hat{C}^{\rm gr}_{x'}\widehat{\sqrt{q_{x'}}}^{1/2}}\ =:\ \hat{h}(x).
\ee 
However, the operator is well defined only in the subspace of $\Hil_{{\rm diff},x}$ corresponding to the positive part of the spectrum of $\widehat{\sqrt{q_{x}}}^{1/2}\hat{C}^{\rm gr}_{x}\widehat{\sqrt{q_{x}}}^{1/2}$. To formulate that condition we need to choose a self-adjoint extension of the operator in the case it is not unique. Denote the resulting subspace of $\Hil_{{\rm diff},x}$ by  $\Hil_{{\rm diff},x+}$. There is a natural averaging map 
\ba 
\eta_M:\ \Hil_{{\rm diff},x} &\rightarrow&  \Hil_{{\rm diff}}\\
\eta_{{\rm Diff}(M,x)}(\psi)\ &\mapsto&\ \eta_{{\rm Diff}(M)}(\psi).
\ea   
The domain of the physical hamiltonian is 
\be 
\Hil_{\rm phys}\ =\ \eta_M(\Hil_{{\rm diff},x+}),
\ee 
and the formula for physical Hamiltonian reads
\be 
\hat{h}_{\rm phys}\ =\ \int d^3x \hat{h}(x)\ =\ \sum_{x\in M} \sqrt{-2\widehat{\sqrt{q_{x}}}^{1/2}\hat{C}^{\rm gr}_{x}\widehat{\sqrt{q_{x}}}^{1/2}}. 
\ee

We remember the anomaly free condition (\ref{[h,h]}) that should be satisfied by our construction. In \cite{MarLew} an extension of the Hilbert space ${\Hil_{\rm phys}}$ is introduced in which the smeared scalar constraint operators 
\be \hat{C}^{\rm gr}(N)\ =\ \int_Md^3x N(x)\hat{C}^{\rm gr}(x) \ee 
are defined and their products
\be \hat{C}^{\rm gr}(N)\hat{C}^{\rm gr}(N') \ee
are contained. It follows from the results of \cite{MarLew} that the smeared constraint operators commute
\be [\hat{C}^{\rm gr}(N),\hat{C}^{\rm gr}(N')]\ =\ 0 \ee     
on a large subspace of the enlarged vector space. It justifies our conjecture, that the condition  
\be [\hat{h}(x),\hat{h}(y)]\ = \ 0\ee 
is a restriction on the ambiguities in the definition of the operators $\hat{h}(x)$, that is on the loop assignment \cite{AshLew1,Thiemm2} and the self-adjoint extensions. 

\section{Concluding remarks, outlook}
We have another quantum model model of gravity involving all the degrees of freedom.
 The model discussed here assumes a vanishing potential for the scalar field that becomes the internal time for the Dirac observables. Neglecting this requirement has the effect that the physical Hamiltonian depends on the internal time $\phi$ as can be seen in equation (\ref{pm}). Non -- conservative Hamiltonians usually increase the intricacy as far as the technical perspective is concerned. Likewise if we use for instance Standard Model matter instead of a scalar field the system will also not deparametrize anymore. Hence,  all the technical simplifications due to deparametrization used in this work are not available any longer. A discussion about which kind of matter Lagrangians induce a deparametrization for General Relativity can be found in \cite{Thiemm3}.
\\
\\
The quantization of this model is complete and every necessary element exists within the framework of LQG. However, there are still ambiguities though, present in the LQG definition of the quantum scalar constraint operator due to its non-polynomial structure.
The only way to understand them and their possible physical meaning is to start applying the model. Before explaining
what the model discussed in this work is good for, let us compare it briefly to the first model that was
completed by Giesel and Thiemann. 
 
\subsection{Comparison with the Brown-Kuchar model applied to LQG} 
The Brown-Kuchar (BK-) model \cite{BroKuch} considers four scalar fields that have the properties of dust and become a dynamically coupled observer, with respect to which the dynamics of the remaining degrees of freedom is formulated. This model was used by Giesel and Thiemann \cite{GieThiemm} and a reduced phase space of gravity coupled to dust was derived. For this purpose the BK-model needed to be extended since the reduced phase space requires also the construction of (classical) Dirac observables with respect to the scalar constraint. The original BK-model is rather the counter part of what is done in this paper because there the vector constraint was reduced classically, whereas for the scalar constraint a quantum condition was formulated.
\\
In the reduced phase space quantization procedure discussed in \cite{GieThiemm} both, the scalar as well as the diffeomorphism constraint, are reduced classically. The Gauss constraint is, as in this paper, solved at the quantum level. This yields to an algebra of observables describing the classical physical phase space. Due to the deparametrization of the scalar constraints, this algebra turns out to be isomorphic to the kinematical one. In contrast to what is done in this paper, a quantization of the observable algebra accesses directly the physical Hilbert space (once also the Gauss constraint is satisfied). Since the kinematical algebra is isomorphic to the physical one, in \cite{GieThiemm} the standard kinematical representation of LQG can also be used for the physical Hilbert space ${\cal H}_{\rm phys}$. Similar to the work in this paper, the generator of the physical dynamics, the so called physical Hamiltonian $h_{\rm phys}$, is invariant under local diffeomorphisms.
In the reduced approach this leads to the requirement, that in order to avoid a quantum anomaly, the operator needs to be invariant under local diffeomorphisms too.  As shown in \cite{ ALMMT} operators being invariant under local diffeomorphisms and defined in the standard (kinematical) LQG representation cannot be graph-changing. This means, that they need to preserve the graph they are acting on, yielding the condition, that the LQG constraint operators  \cite{Thiemm1, AshLew1} entering the physical Hamiltonian $\hat{h}_{\rm phys}$ in \cite{GieThiemm} need to be quantized in a graph-preserving way.
As we explained above,  LQG is glued from the Hilbert spaces corresponding to all possible graphs. The original LQG scalar constraint  operator does not preserve those graph Hilbert spaces. In the   model of \cite{GieThiemm} the physical Hamiltonian must preserve each graph  Hilbert space. In the consequence,  the constraint operator has to be suitably  redefined in \cite{GieThiemm} when the standard (kinematical) LQG representation is used for ${\cal H}_{\rm phys}$. The paper \cite{GieThiemm} also discusses the quantization of the reduced model in the framework of Algebraic Quantum Gravity \cite{AQGI}, where a different representation is used, namely von Neumann's infinite tensor product representation.  The quantum dynamics is not defined on embedded graphs but on abstract ones. carrying only combinatorial information. In this framework only the infinite combinatorial graph, that the theory is defined on and that acts like an abstract lattice, needs to be preserved by $\hat{h}_{\rm phys}$, whereas any possible subgraph of this does not. In the case of the model presented in this paper here, the graph Hilbert spaces are not preserved and they evolve in the emergent time.  
                       
\subsection{Application of this model}
Our model can be used to verify the properties of quantum space-time we expect
after learning the lessons from LQC and QFT in curved spacetime. 

In the LQC models of the homogeneous massless scalar field coupled to gravity, Big Bang turns out to be replaced by Big Bounce, as the result of the quantum  gravity effects. Now, with our model, we can consider the same system of fields from the point of view of the full theory, without the symmetry reduction. Similarly, we can also consider the quantum gravitational collapse, quantum black holes, theory entropy.
All those cases are manageable within our model, and the only difficulty is of technical nature.  
Also the Hawking radiation and black hole evaporation process expected from the
theory of quantum fields on the classical black hole background are in the range
of our model.  
The next step to obtain progress in this direction is the construction of semiclassical states for full LQG, which are preserved under quantum dynamics generated by the physical Hamiltonian on appropriate time scales.
\\
In conclusion, our paper opens the door to understanding the properties of quantum spacetime from the point of view of the full quantum gravity.

\begin{acknowledgements}
We thank Abhay Ashtekar, Benjamin Bahr, Frank Hellmann,  Youngge Ma, Carlo Rovelli and Thomas Thiemann for their comments. K. Giesel thanks the ESF sponsored network 'Quantum Geometry and Quantum Gravity' for a short visit grant to collaborate with the Warsaw group. Part of this research was supported by the DFG cluster of excellence 'Origin and Structure of the Universe'. The remaining authors  were  partially supported by:  (i) the grants N N202 104838, N N202 287538 and 182/N-QGG/2008/0 (PMN) of Polish  Ministerstwo Nauki i Szkolnictwa Wyzszego; (ii) the grant Mistrz of the Fundation for the Polish Science. J. Lewandowski also acknowledges the US National Science Foundation (NSF) grant PHY-0456913.
\end{acknowledgements}

\end{document}